\documentclass[8.5pt,twoside,twocolumn]{article}
\usepackage[super,sort&compress,comma]{natbib} 
\usepackage{mhchem}
\usepackage{amsmath}
\usepackage{amssymb}
\usepackage{cmap}
\usepackage{tikz}
\usepackage{sectsty}
\usepackage{balance} 
\usepackage{xcolor}
\usepackage{graphicx}
\usepackage{lastpage}
\usepackage[format=plain,justification=raggedright,singlelinecheck=false,font=small,labelfont=bf,labelsep=space]{caption} 
\usepackage{fancyhdr}
\newcommand{\tikzcircle}[2][red,fill=red]{\tikz[baseline=-0.5ex]\draw[#1,radius=#2] (0,0) circle ;}%

\begin{document}

\twocolumn[
  \begin{@twocolumnfalse}
\noindent\LARGE{\textbf{Ionic effects in self-propelled Pt-coated Janus swimmers}}
\vspace{0.6cm}

\noindent\large{\textbf{Aidan Brown$^{\ast}$\textit{$^{a}$} and Wilson Poon\textit{$^{a}$}}}\vspace{0.5cm}

\noindent\textit{\small{\textbf{Received Xth XXXXXXXXXX 20XX, Accepted Xth XXXXXXXXX 20XX\newline
First published on the web Xth XXXXXXXXXX 200X}}}

\noindent \textbf{\small{DOI: 10.1039/b000000x}}
\vspace{0.6cm}

\noindent \normalsize{Colloidal particles partially coated with platinum and dispersed in \ce{H2O2} solution are often used as model self-propelled colloids. Most current data suggest that neutral self-diffusiophoresis propels these particles. However, several studies have shown strong ionic effects in this and related systems, such as a reduction of propulsion speed by salt. We investigate these ionic effects in Pt-coated polystyrene colloids, and find here that the direction of propulsion can be reversed by addition of an ionic surfactant, and that although adding pH neutral salts reduces the propulsion speed, adding the strong base NaOH has little effect. We use these data, as well as measured reaction rates, to argue against propulsion by either neutral or ionic self-diffusiophoresis, and suggest instead that the particle's propulsion mechanism may in fact bear close resemblance to that operative in bimetallic swimmers.}

\vspace{0.5cm}
 \end{@twocolumnfalse}
  ]

\footnotetext{\textit{$^{a}$~SUPA, School of Physics and Astronomy, University of Edinburgh, JCMB Kings Buildings, Edinburgh EH9 3JZ, United Kingdom; E-mail: abrown20@staffmail.ed.ac.uk}}

\section{Introduction}
A current frontier in physics is the study of intrinsically non-equilibrium active matter \cite{Sriram}, including suspensions of self-propelled colloids \cite{PoonVarenna2}. These active colloids show a variety of novel collective phenomena, whose systematic study may pave the way for new theories of non-equilibrium statistical physics. To meet the challenge of understanding such phenomena, experimental data from well-characterised model systems are required. 

An effective model system of self-propelled colloids should ideally have well-defined size, shape and inter-particle interaction, while fuel consumption and waste production rates should permit 3D experiments at high volume fraction. Furthermore, the propulsion mechanism should be well understood, particularly as the flow~\cite{spagnolie12} and concentration~\cite{theurkauff12} fields associated with the propulsion will themselves modify interparticle interactions.

One candidate model system consists of polystyrene (PS) colloids half-coated with platinum (Pt) and suspended in hydrogen peroxide (\ce{H2O2}). These have been used to study active diffusion \cite{howse07}, sedimentation  \cite{BocquetSediment}, and phase separation \cite{palacci10}.

\begin{figure}[t]
  \centering 
\includegraphics[width=6cm, bb = 0 0 475 525]{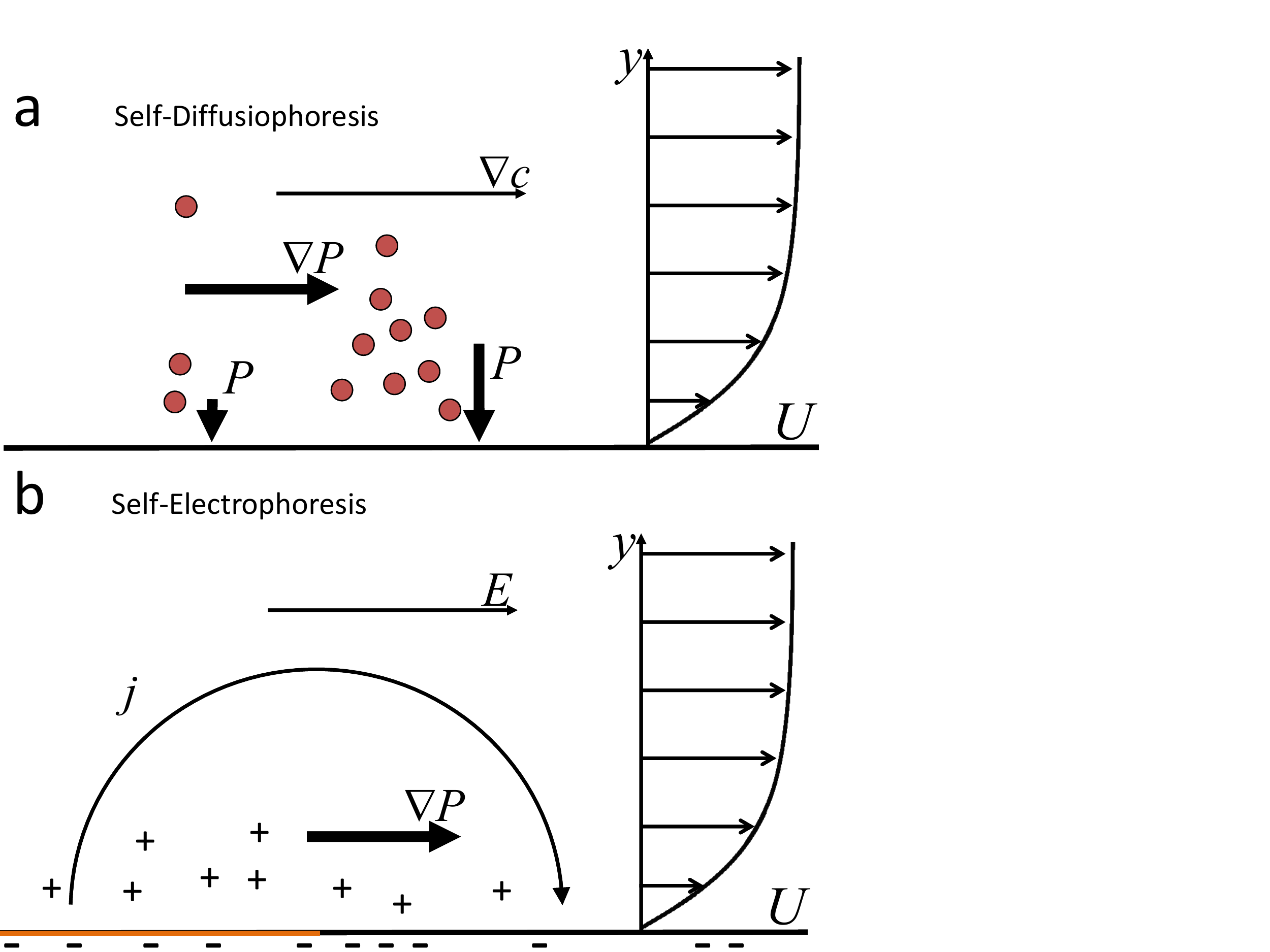}
  \caption{a) Schematic of self-diffusiophoresis~\cite{ajdari06}. Chemical reaction on the particle surface generates a concentration gradient $\nabla c$, indicated by the number of solute molecules (red circles) at the particle surface. A potential interaction  between surface and molecules generates a tangential pressure gradient $\nabla P$, which drives tangential fluid flow.  b) Self-electrophoresis. Here a chemical reaction produces an ionic current $j$, which generates an electric field $E$. Since the particle is charged (-ve here), there is an increased concentration of counterions (+ve) in the Debye layer. The electric field acting on these ions again drives fluid flow. The current is typically produced by the difference in electron affinity between two different metals (i.e. Au-Pt~\cite{paxton06}), but could also be produced by more general assymetries.}
\label{schematic}
\end{figure}

\begin{figure}[t]
  \centering 
\includegraphics[width=8cm, bb = 0 175 475 525]{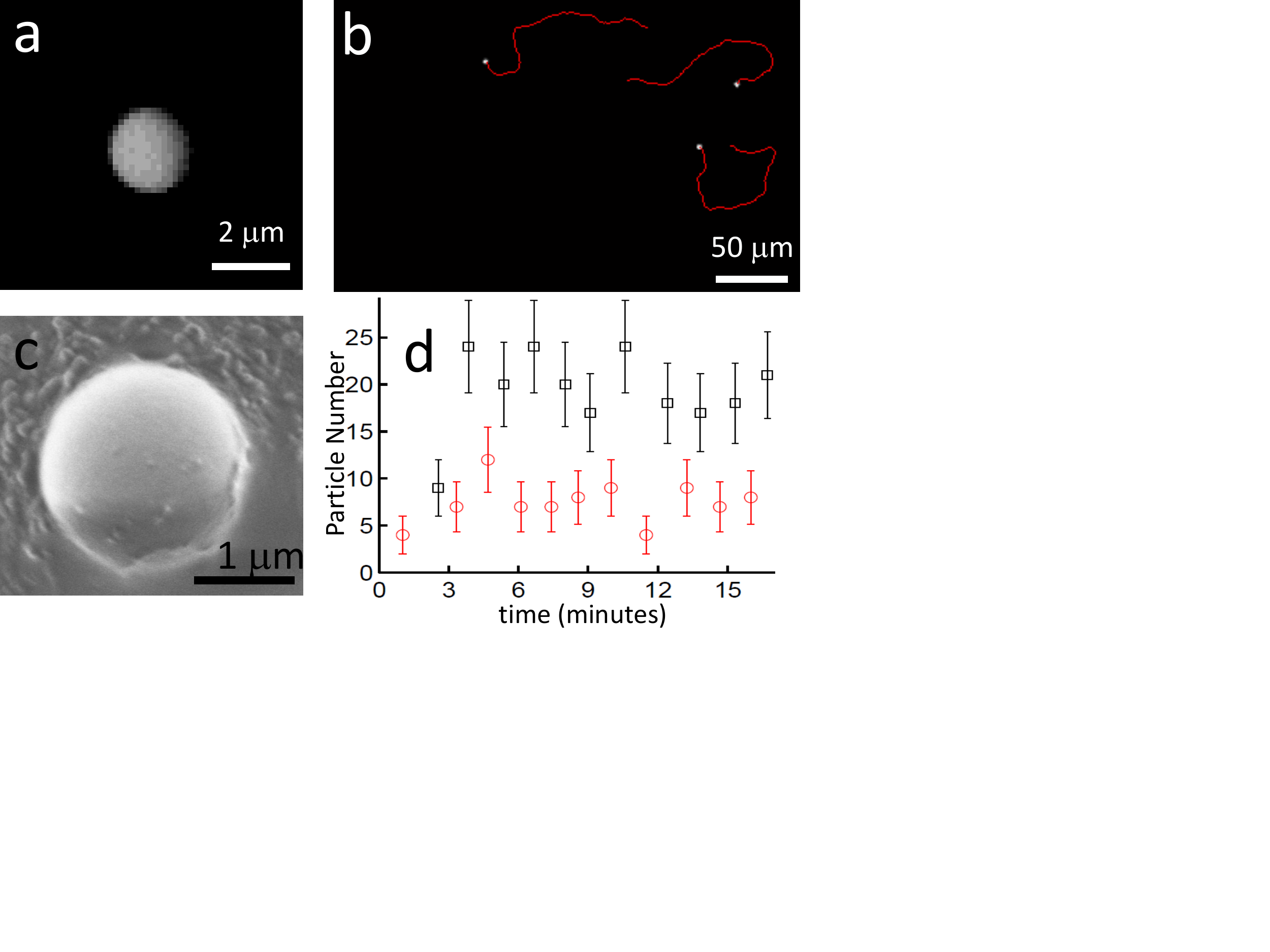}
  \caption{a) Epifluorescence image of a Janus particle at high magnification: the darker region of the particle is coated in platinum. b) Typical tracked video in $\mathrm{10\%}$ \ce{H2O2} c)  Scanning electron micrograph of typical PS-Pt Janus particle (Hitachi SEM at 1 kV, without further coating). The lighter region is Pt. d) Number of particles on the upper (black squares) and lower (red circles) surfaces of a capillary as a function of time after particles were placed in the capillary. The capillary depth is 400 $\mathrm{\mu m}$, and the field of view is 800 $\times$ 800 $\mathrm{\mu m}$. }
\label{figure1a}
\end{figure}

These Pt-PS Janus colloids originated from predictions~\cite{golestanian05} that colloids with an asymmetric catalytic coating should propel themselves via self-diffusiophoresis in a substrate solution. Diffusiophoresis is the propulsion of a particle in a solute gradient, due to interactions between the solute and the particle surface~\cite{anderson84}. In self-diffusiophoresis, this gradient is generated by reactions on the particle's surface. For Pt-PS Janus particles, it has been proposed that the \ce{O2} and \ce{H2O2} gradients produced by the decomposition of \ce{H2O2} on Pt directly drives this self-diffusiophoresis~\cite{howse07}, Fig.~\ref{schematic}(a). Successful prediction of how \ce{H2O2} concentration ([\ce{H2O2}]) \cite{howse07} and particle radius \cite{ebbens12} control propulsion speed support this picture.

Nevertheless, there is also some evidence that this simple propulsion model cannot tell the whole story of heterogeneous particles swimming in \ce{H2O2}. For example, catalytic Janus swimmers have been observed moving both towards their catalytic~\cite{palacci13, valadares10}, and their non-catalytic~\cite{ebbens11} face; and salt~\cite{palacci13} and anionic surfactants~\cite{gibbs09} have both been observed to reduce the the swimming speed of Janus swimmers. Such observations are not predicted by the simple model of self-diffusiophoresis in gradient of neutral species, and are more suggestive of the ionic propulsion mechanisms of bimetallic swimmers~\cite{paxton04}, where motion towards either pole~\cite{wang06} and speed reduction with increasing salt concentration~\cite{paxton06} are both well understood as integral features of the propulsion mechanism. 

In this paper, we measure the effect of several ionic species, a cationic surfactant, two neutral salts and a basic salt, on the swimming speed of Pt-PS Janus swimmers. We confirm that the neutral salts do decrease the propulsion speed of the particles, and show by direct measurement that this is only due in a small part to a concomitant decrease in reaction rate. We find, however, that NaOH, which alters the solution's pH, produces no significant speed reduction. We also find that adding the cationic surfactant CTAB not only reduces, but eventually reverses the propulsion velocity. 

Our chief objective is to argue that it is difficult to reconcile these ionic effects with propulsion due to diffusiophoresis driven either by the neutral species (\ce{O2} and \ce{H2O2}) taking part in the Pt-catalysed decomposition of \ce{H2O2}, or by ions generated by the partial dissociation of \ce{H2O2}. Based on this conclusion, we suggest that the propulsion mechanism could be the same as that of bimetallic swimmers, where an ionic current generates motion via self electrophoresis~\cite{paxton06}, Fig.~\ref{schematic}(b). We point out that such a current could be produced by general asymmetries in the particle surface, and does not necessarily require a bimetallic particle. Showing whether this mechanism is in fact operative will require future, detailed experiments and theoretical modelling.

\section{Materials and Methods}

Janus particles were prepared by sputtering approximately 5 nm Pt onto sulfonated fluorescent 2~$\mathrm{\mu m}$ diameter PS particles (Invitrogen), cleaned 3 times in deionised water, and deposited on glass coverslips at sub-monolayer concentrations. Janus particles were resuspended by sonicating for 20 minutes in deionized water.  Fig.~\ref{figure1a}(c) shows an SEM image of a typical Janus particle. This method produced mainly isolated particles, but a small fraction of particles remained bound in 2-3 particle clusters after sonication.

Samples containing $\mathrm{10^{-7}}$ v/v Janus particles in $\mathrm{10\%}$ aqueous \ce{H2O2} (Acros) and varying concentrations of KBr, NaCl, NaOH, or cetyl-trimethyl-ammonium bromide (CTAB) (all $\geq 97\%$ purity), were placed in $0.4\times 8 \times50$~mm glass capillaries (Vitrocom). Particle motion on the upper or lower capillary surface was recorded in $20\times$ epifluorescence with a CCD camera (Eosens, Mikrotron; 400 frames at 20 frames per second) and tracked by standard algorithms~\cite{crocker96}. Fifty tracks from 2 capillaries were averaged for each condition. Particle clusters were not tracked. The ballistic speed $v$, and translational diffusivity $D$, were obtained by fitting to the first 3 frames of the particles' mean squared displacement~\cite{howse07}, and the polarity of motion was deduced from the shadowing effect of the Pt coating~\cite{ebbens11}, Fig.\ref{figure1a}(a). The particles quickly reach the capillary surfaces, Fig.~\ref{figure1a}(d), so all quantitative measurements were performed at the capillary surface, although bulk observations confirmed that phenomenology remained unchanged away from the surface, as discussed below.

Reaction rates of $\ce{H2O2}$ on Janus particles were measured by weighing polystyrene fluorimetry cuvettes (Fisher Scientific) containing 3 ml of 10$\%$ \ce{H2O2} with and without 0.5 mM or 1 mM NaCl, or 1 mM NaOH, with $1 \pm 0.5 \times 10^{-5}$ v/v Janus particles repeatedly for 2 hours. The higher volume fraction required for reaction rate determination required deposition of colloids in an almost complete monolayer. Consequently, following sonication, there was a substantial fraction of large ($>$5 particle) aggregates. This accounts for the large uncertainty in volume fraction, which was determined by counting approximately 1000 particles at known dilution. 

Cuvettes were wrapped in aluminium foil to keep out the light, since it was found that under room lighting, cuvettes of 10$\%$ \ce{H2O2} without Janus particles displayed a significant background mass loss, probably caused by photocatalysis of the hydrogen peroxide decomposition. In darkness, 10$\%$ \ce{H2O2} without Janus particles showed mass loss of less than 1 mg over two hours, much lower than the mass loss with Janus particles, at approximately 80 mg over two hours. 

Sonication may release small flakes of Pt from the film sputtered onto the glass coverslip, contributing to measured reaction rates. No such flakes were observed at optical resolution in our samples. For a hexagonally packed layer, the area fraction of the interstices is approximately 10$\%$, whereas the Janus particles, of radius $R$ have an area of $2\pi R^2$ of Pt per $\pi R^2$ of glass which they cover. One would therefore expect, assuming all Pt flakes separate from the surface, and that both sides of the Pt contribute to the reaction, that these Pt flakes contribute at most $10\%$ of the measured reaction rates.

The point of zero charge of polystyrene particles in CTAB solution was determined by measuring the electrophoretic mobility of polystyrene particles ($10^{-3}$ v/v) in solutions of varying concentration of CTAB in 10$\%~\ce{H2O2}$ using a Malvern Nanosizer. The capillary cells used platinum electrodes, and production of oxygen bubbles on the electrodes was avoided by using low voltages ($<40$ V), and monomodal analysis, which does not use prolonged DC voltage sweeps. The Nanosizer was also used to measure the conductivity of $10\%$ \ce{H2O2}.

\section{Results}

\begin{figure}[t]
  \centering 
\hspace*{1cm} \includegraphics[width=11 cm, bb = 150 525 550 750]{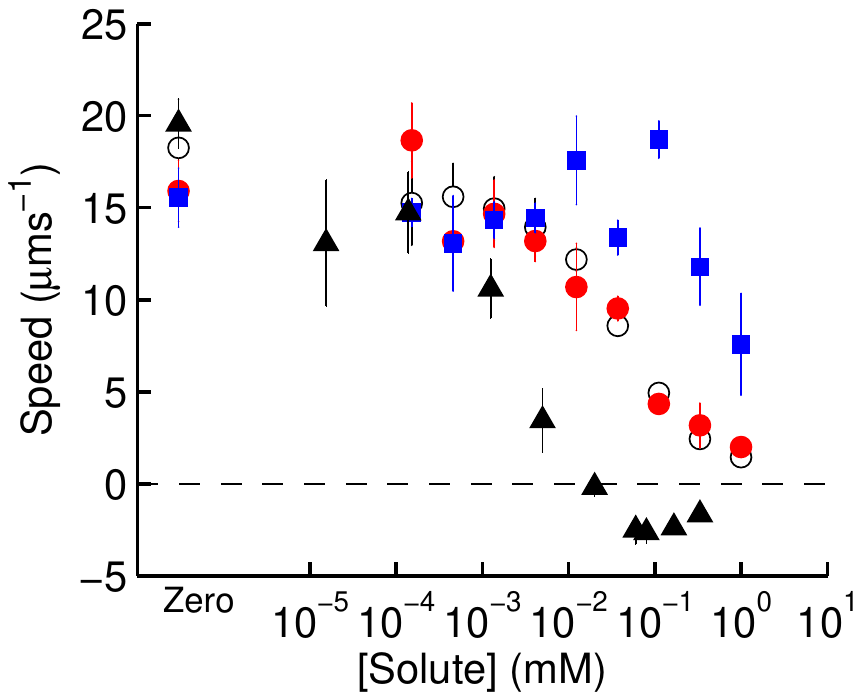}
\hspace*{1cm}\includegraphics[ width=11 cm, bb =150 575 550 725]{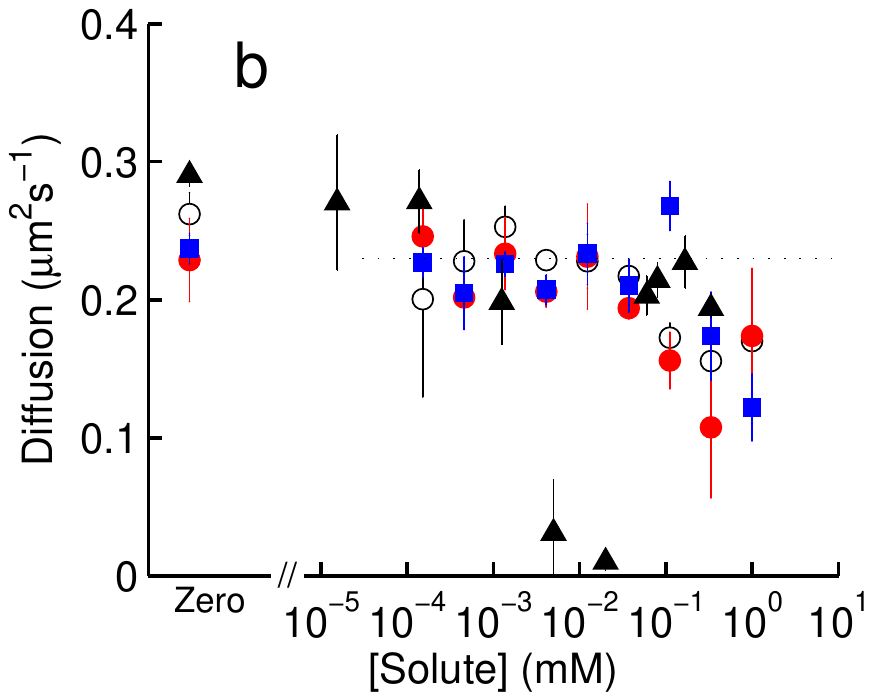}
  \caption{(a) Speed $v$ as a function of concentration of CTAB ($\blacktriangle$), KBr ( \tikzcircle{2pt}) , NaCl($\circ$), and NaOH (\textcolor{blue}{$\blacksquare$}). Positive $v$ indicates propulsion towards PS. Leftmost points correspond to zero concentration. (b) Translational diffusion $D$ for the same data. The dotted horizontal line corresponds to the predicted bulk diffusion constant $D=k_BT/(6\pi\eta R)=0.23\mathrm{~\mu m^2s^{-1}}$.}
\label{All figures plot}
\end{figure}

Particles reach the capillary surface within a few minutes, Fig.~\ref{figure1a}(d). We find that as-prepared Janus particles in \ce{H2O2} swim towards their PS face, and predominantly go towards the upper surface, whereas with sufficient CTAB, they swim towards their Pt face and predominantly go towards the lower surface. This is consistent with the previously reported gravitactic nature of these particles~\cite{campbell13}, where buoyancy results in a preference for the heavier Pt hemisphere to face downwards. Once particles reach the capillary surface, they remain there indefinitely, moving parallel to the surface, even if the capillary is inverted. 

In 7 independently manufactured batches, particles always moved consistently towards their PS face in $\mathrm{10\%~ H_2O_2}$ only, at an average speed of $\langle v\rangle=\mathrm{11~\mu m s^{-1}}$ with a standard deviation of $\sigma_v=\mathrm{6~\mu m s^{-1}}$. Figure~\ref{All figures plot}a shows speed and direction measured for a representative batch of Janus particles, in 10$\%$ \ce{H2O2} and a range of ionic solutes. The behaviour seen here is qualitatively similar in all other Janus samples observed, but, as stated above, the overall speed varies from sample to sample. 

We find that $v$ decreases with increasing [CTAB], and that the direction of motion, initially towards PS, reverses at $\mathrm{[CTAB] \sim20 \mu M}$. In order to check that this reversal is not a wall effect, we observed Janus particles moving in the bulk around the point where CTAB reverses the direction of motion. We find that the direction reversal occurs in the bulk at [CTAB]=$19\pm 2\mathrm{~\mu M}$. This value was obtained by counting particles visibly moving towards the Pt or PS face over a range of [CTAB] between 5 and 50 $\mathrm{\mu M}$, observing 10 particles at each concentration.

It is known that CTAB adsorbs onto anionic PS particles first to neutralise and then reverse the charge~\cite{ottewill}. If CTAB causes the direction reversal through a charge effect, then we would naively expect the charge reversal point to correspond to the direction reversal. We found that initially negatively charged uncoated PS particles  in $10\%~\ce{H2O2}$ become positive at $\mathrm{[CTAB] = 3.9 \pm 0.1 \times 10^{-3}~mM}$, approximately 20$\%$ of the value required for direction reversal of the Pt-Janus swimmers. However, we do not know whether CTAB also adsorbs onto Pt, since low particle yield prevented similar electrophoretic measurements on Pt Janus particles. The diffusion constant of the particles remained approximately constant as a function of [CTAB] apart from between 5 and 20 $\mu$M, where the particles appear to be stuck to the capillary surface. Presumably, this is because the glass surface has an opposite charge to the Pt-Janus swimmers within this concentration range.

With neutral salts (NaCl and KBr), we observe a reduction in $v$, Fig.~\ref{All figures plot}(a), but no reversal in swimming direction. This fall in speed is again not simply a surface effect, since visual observation confirmed a steady decrease in visible particle speed in the bulk, until there was no visible motion of particles in the bulk at 1 mM salt. In fact, particles in the bulk remained there for over an hour in 1 mM salt, while without salt they reached the top surface within a few minutes. In addition, if the reduction in particle speed were caused by, for example, a decrease in Debye length leading to increased frictional drag due to closer contact with the surface, one would expect the particles' diffusivity to decrease by a similar amount. Instead, we find that particle diffusivity falls by approximately $30\%$ up to 1 mM salt, and remains close to the predicted bulk value (dotted line in Fig.~\ref{All figures plot}(b)). Adding up to 1 mM NaOH produced no noticeable reduction in particle speed, Fig.~\ref{All figures plot}(a), which again is reflected in bulk observations.

We also prepared Janus particles with thicker, 10 nm Pt coatings. These particles displayed irreproducibility in the direction of motion, even in 10$\%$ \ce{H2O2} only, with some samples showing propulsion towards the Pt face, some towards the PS face, and some propulsion in both directions. In all cases, a reduction in speed was observed in increasing salt concentration, as in Fig.~\ref{All figures plot}. Additionally, the direction of motion was reversed upon addition of approximately 20 $\mathrm{\mu M}$ CTAB, with those particles which were initially moving towards their Pt face, moving towards their PS face, entirely reversing the behaviour shown in Fig.~\ref{All figures plot}(a). Because of the irreproducibility of these particles' behaviour, they were not studied systematically, but the qualitative implications of these observations are discussed below.

Reaction rates were obtained by fitting the time-dependent mass to exponential decays, Fig.~\ref{spiral}, which were averaged over at least 3 cuvettes for each sample condition. In $10\%~\ce{H2O2}$, we calculate that each particle consumes \ce{H2O2} molecules at rate $\Gamma_0= 8 \pm 4 \times 10^{10}\mathrm{~s^{-1}}$, which is comparable to Au-Pt swimmers~\cite{wang13, paxton04}. With 0.5 mM NaCl, the relative rate falls to $\Gamma/\Gamma_0= 0.83 \pm 0.02$, with 1 mM NaCl, to $\Gamma/\Gamma_0= 0.58 \pm 0.06$, and with 1 mM NaOH to $\Gamma/\Gamma_0= 0.44 \pm 0.04$. Note that this relative uncertainty is much less than the absolute uncertainty, which mainly comes from the uncertainty in volume fraction arising form the presence of clusters. 

\section{Discussion}

We now proceed to argue that our results, Figs.~\ref{All figures plot} and \ref{spiral}, are not predicted by, and are largely incompatible with, both neutral and ionic diffusiophoresis. Under the heading of neutral diffusiophoresis (section \ref{sec:neutral}), we successively consider diffusiophoresis due to excluded volume (section \ref{sec:excluded}), Van der Waals (VdW) and hydrophobic interactions (section \ref{sec:VdW}) between neutral species and the particle surface. Such a mechanism is shown to be incompatible with our observation of  salt effects (section \ref{sec:salt}), and with our measured reaction rates (section \ref{sec:rates}). Next, we argue that self diffusiophoresis in a gradient of ions generated by the spontaneous dissociation of \ce{H2O2} cannot explain our observation either (section \ref{sec:ionicdiff}), before offering a speculative mechanism that is compatible with, but not necessarily implied by, our data (section \ref{sec:electro}). Detailed calculations behind some of the arguments in sections \ref{sec:neutral} and \ref{sec:ionicdiff} are given in two appendices.

\subsection{Neutral Self-Diffusiophoresis} \label{sec:neutral}

\begin{figure}
\centering
\hspace*{1cm}\includegraphics[width=10 cm, bb = 150 575 550 750]{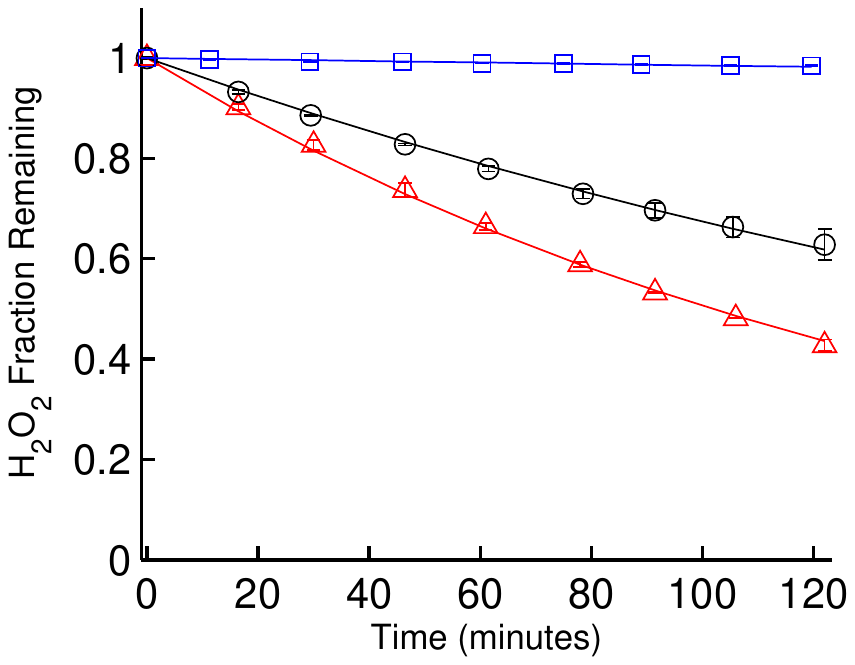}
\caption{Fractional mass loss of \ce{H2O2} over time from $10^{-5}$ v/v Janus particles in $10\%$ \ce{H2O2}, with ($\circ$) or without (\textcolor{red}{$\triangle$}) 1~mM~NaCl. The control (\textcolor{blue}{$\square$}) is $10\%$ \ce{H2O2} without Janus particles. Solid lines are exponential fits to the data.}
\label{spiral}
\end{figure}

The prevailing explanation for the propulsion of catalytic swimmers is neutral self-diffusiophoresis~\cite{golestanian05,howse07,ebbens12}. In this model, the direct interactions between the neutral molecules involved in the reaction:
\begin{eqnarray*}
	\ce{2H2O2  <=>  2H2O + O2} \,\, 
\end{eqnarray*}
and the particle surface, generate tangential pressure gradients which drive fluid flow along the particle surface.  In the reaction rate dominated regime~\cite{ebbens12}:
\begin{eqnarray}
	 v \,=\, -\frac{\Gamma k_BT\alpha_{\rm eff}}{8\pi D_\ce{H2O2} R^2\eta}                 \,, \label{neutral speed}
\end{eqnarray}
where $k_BT$ is the thermal energy, $\eta$ the viscosity, $R$ the particle radius and $D_\ce{H2O2}$ the \ce{H2O2} diffusivity. $\alpha_{\rm eff}$, defined in Appendix I, characterises the interaction between the product of catalytic decomposition and the particle surface. $\alpha_{\rm eff}$ can be used to define the characteristic lengthscale of the interaction $\lambda_{\rm eff}$, called the Derjaguin length, through $\lambda_{\rm eff}^2=|\alpha_{\rm eff}|$~\cite{ebbens12}. $\alpha_{\rm eff}$ is positive if the reaction products interact more favourably with the particle surface than the reactants do, and in this case the particle will move towards the Pt surface ($v<0$). Hence, the initial observed direction of motion of the particles towards PS implies that, in this case, the reactants (\ce{H2O2}) interact more favourably with the particle surface than the products (\ce{O2}). 

Taking our average speed in $10\%$ \ce{H2O2}, $v_0=\mathrm{17\pm1.5~\mu ms^{-1}}$, and using $T = 300$~K, $\eta = 10^{-3}$ and $D_\ce{H2O2}=6\times10^{-10}~\mathrm{m^2s^{-1}}$ for the \ce{H2O2} diffusivity~\cite{hall98}, we obtain $\mathrm{\alpha_{eff}=0.8\pm 0.4~nm^2}$. This is 2 orders of magnitude higher than the previously-estimated value for Pt-PS Janus swimmers~\cite{ebbens12}. In the previous work, the reaction rate $\Gamma$ in Eq.~\ref{neutral speed} was estimated from the assumption of a diffusion limited reaction, for which the rate is $\Gamma_D\sim2\pi D_\ce{H2O2}[\ce{H2O2}] R =7\times10^{12} \mathrm{~s^{-1}}$. Our measured reaction rate at zero salt, $\Gamma_0$, is approximately $1\%$ of $\Gamma_D$, implying that Pt-PS particles function well below the diffusion limit. Using our measured reaction rate, rather than an estimated value, accounts for the significantly revised value of $\lambda_{\rm eff}$. 

If diffusiophoresis due to some neutral species is the dominant propulsive mechanism, our experimental data place stringent constraints upon the interaction between these species and the particle surface. This interaction has to have an effective length scale $\lambda_{\rm eff}$ of order 1 nm; its nature must be such that Eq.~\ref{neutral speed} initially predicts propulsion away from the Pt, but be capable of reversal with the addition of CTAB; and its strength must decrease with increasing ionic concentration, though not when NaOH is used instead of a pH-neutral salt. 

Note that the decrease in speed with concentration of neutral salts, Fig.~\ref{All figures plot}, does not have to be due entirely to a decrease in interaction strength, because we have seen that the reaction rate decreases at high salt, Fig.~\ref{spiral}. However, this still leaves a factor of approximately 5 in speed reduction between 0 and 1 mM salt to be accounted for by a decrease in interaction strength.

\subsubsection{Excluded Volume Interactions} \label{sec:excluded}

Our results certainly rule out the simplest interaction proposed in the literature, which is excluded volume~\cite{brady11, cordovafigueroa08}. First, and most basic, is the fact that excluded volume interaction should return a $\lambda_{\rm eff} \sim \mathrm{\AA}$, i.e. of the order of a molecular radius; instead we find $\lambda_{\rm eff} \sim \mathrm{nm}$. Secondly, salt should have no effect beyond changing reaction rate, and we have seen that this is not true in our case. Finally, excluded volume interactions cannot explain the observation of velocity reversal. 

We note, in passing, that the observation of speed reversal also rules out models of swimming which are propelled purely by invisible nanobubbles~\cite{gibbs09}, since bubbles should only be produced from the Pt coated side. Note that bubble-propelled swimmers do exist~\cite{solovev09, gao11, manjare12} but these tend to be larger, and leave a visible train of bubbles, which are not observed in the current system.

\subsubsection{Van der Waals and Hydrophobic Interactions} \label{sec:VdW}

Ruling out neutral diffusiophoresis due to interactions more complicated than excluded volume is less straightforward. The relevant length scales are small enough that molecular details of the liquid-particle interface matter. Thus, continuum models~\cite{anderson89, golestanian05} are probably no longer applicable, and discrete molecular perspectives~\cite{brady11} may be necessary. Moreover, experimentation may never be able to offer conclusive proof that specific mechanisms are not operative. What we will show, however, is that peculiar circumstances are required to render neutral diffusiophoresis compatible with our data. The relative ease with which Janus swimmers in \ce{H2O2} can be prepared argues strongly against the operation of such `special circumstances'.

When metals are present, VdW interactions are strong~\cite{israelachvili11}. Thus, we first consider whether VdW interactions can power neutral diffusiophoresis in our particles. The VdW interaction notoriously diverges at contact. However, our Pt surface is probably covered with a layer of bound water, oxide or hydroxide groups~\cite{katsounaros12, sitta14, climent06, bianchi62}, preventing direct \ce{O2}-surface contact. In Appendix I, we estimate a value of $\approx 0.2 \,\mathrm{\mu m s^{-1}}$ for the expected diffusiophoretic speed under a VdW interaction between \ce{O2}, \ce{H2O2} and the Pt surface, assuming that there is an immobile layer, one water diameter (0.25~nm)~\cite{israelachvili83} thick. This estimated speed is $\sim 50$ times smaller than observed, Fig.~\ref{All figures plot}. The normal assumption in modelling diffusiophoresis is that there is a strict no-slip boundary condition at the particle liquid interface. However, if there is instead a finite slip length at the surface, as occurs for very smooth hydrophobic surfaces~\cite{cottinbizonne05},  the diffusiophoretic velocity can be amplified by a factor of $1+b/L$, where $b$ is the hydrodynamic slip length, and $L$ is a typical lengthscale of the interaction\cite{ajdari06}. Thus, a slip length of order 80 molecular diameters (or approximately 8 nm) could conceivably increase our estimate of $\sim 0.5\,\mu$ms$^{-1}$ to the observed range of speeds. While such a slip length is plausible~\cite{cottinbizonne05, sendner09}, it would appear to be incompatible with the presence of bound \ce{OH} or \ce{H2O} groups at the Pt surface, and achieving such slip lengths usually requires a carefully prepared, smooth, hydrophobic surface.

The hydrophobic interaction and hydration forces may be larger than the VdW interaction~\cite{israelachvili11, meyer06}. However, these entropic interactions remain poorly understood and poorly characterised; in particular, measurements of hydrophobic interactions between surfaces may not translate to molecular interactions~\cite{meyer06, israelachvili11}.

\subsubsection{Propulsion Direction Reversal} \label{sec:reverse}

The reversal of propulsion by CTAB could be explained simply in a neutral diffusiophoretic theory if there are several distinct, competing types of interaction, such as VdW and hydrophobic interactions, or if a single type of interaction is capable of reversing sign. For example, we would expect CTAB to modify the hydrophobicity of both the PS and Pt surfaces, and this could reverse the sign of the hydrophobic interaction. If the reversal in direction is indeed due to a change in hydrophobicity, other surfactants would also be expected to modify the propulsion speed or direction. The evidence on this point is inconclusive. The anionic surfactant sodium dodecylsulfate (SDS) has previously been found to reduce the speed of Pt-\ce{SiO2} swimmers in \ce{H2O2}~\cite{gibbs09}. However, the high concentrations used (of order 1 mM~\cite{mysels86}) means that the SDS molecules remaining in solution probably exert their effect as simple salts rather than as surfactants, and there may be additional effects of the surfactant on reaction rate. It has also been observed that the hydrophobicity of the uncoated face of the Janus swimmer can modify the swimming speed, with more hydrophobic swimmers moving up to 50$\%$ faster than hydrophilic swimmers~\cite{manjare14}, but there it was found there that the variation in swimming speed was correlated with an increased reaction rate on the hydrophobic surface. On the other hand, the reversed propulsion direction which we sometimes see in as-prepared Janus particles with thicker Pt coatings is difficult to account for by neutral diffusiophoresis.

\subsubsection{Salt Effects} \label{sec:salt}

The generic effect of salt on electrostatic interactions, including the static part of the VdW interaction, is to introduce a screening factor $\exp{(-D/\lambda_D)}$, where $D$ is the distance between the molecule and the surface~\cite{israelachvili11}. However, the typical distances involved in any neutral interactions are much smaller than the Debye length in our system ($\approx 10$~nm for 1 mM salt), so varying $\lambda_D$ should have very little effect, contrary to our data and others' observations~\cite{palacci13}. In addition, the fluctuating part of the VdW interaction, which dominates for interaction across water, is not screened at all by salt~\cite{israelachvili11}. The effect of salt on the hydrophobic interaction is unclear. Both strenthening and weakening of the interaction by salt have been observed, but many of these observations have been attributed to electrostatic artefacts~\cite{meyer06}. Finally, our observation that the type of salt matters (NaOH produces little effect), indicates that additional pH dependent behaviour is required, which is not predicted.

\subsubsection{Reaction Rates} \label{sec:rates}

Independent of any of the above considerations, our measured reaction rates provide evidence agains neutral diffusiophoresis. In previous work~\cite{ebbens12, campbell13, manjare14}, a propulsion speed inversely proportional to particle radius has been observed for Pt-PS Janus swimmers similar to ours. This radius dependence was ascribed in the original work~\cite{ebbens12} to assumed diffusion limited fuel consumption, and this assumption was in fact used to estimate the reaction rate there. However, direct measurements show that our particles are in the reaction rate limited regime, with reaction rates 2 orders of magnitude lower than the diffusion limit. Since the previous study~\cite{ebbens12} shows propulsion at similar speeds (10 $\mathrm{\mu ms^{-1}}$ for 1 $\mathrm{\mu m}$ radius swimmers in 10\% \ce{H2O2}\cite{ebbens12}) to our work, we may reasonably assume that the fuel consumption rates are also similar. We therefore conclude that, in fact, the previous work was also performed under reaction rate limited conditions. If so, this implies a contradiction with the neutral diffusiophoretic model~\cite{ebbens12}, which predicts a $1/R$ scaling of speed if and only if the reaction is diffusion limited.

\subsubsection{Summary}

Thus, our data impose constraints such as a $\sim 1$~nm length scale of the interaction, a large reduction in speed in 1~mM pH-neutral salt, and the reaction-rate-limited consumption of \ce{H2O2}, which together add up to a strong case against self-diffusiophoretic propulsion due to the interaction of a neutral species with the particle surface. Some very special conditions, such as a large slip length, or particular combinations of hydrophobic interactions with multiple neutral species, may conceivably make this mechanism work in isolated instances. However, the ubiquity of self propulsion in Janus particles and related systems relying on the catalysed decomposition of \ce{H2O2} means that neutral diffusiophoresis is very unlikely to be the generic operative mechanism.  Furthermore, attributing the propulsion of Pt-PS Janus particles to neutral diffusiophoresis driven by hydrophobic or hydration forces of unknown origin and strength, is rather unsatisfactory. 

\subsection{Ionic Self-Diffusiophoresis} \label{sec:ionicdiff}

The strong effect of pH-neutral salts, Fig.~\ref{spiral}, suggests ionic self diffusiophoresis as an alternative mechanism. Although neither ionic reactants nor products feature in \ce{H2O2} decomposition, \ce{H2O2} undergoes ionic disocciation via~\cite{kosinski07}:
\begin{eqnarray}
	\ce{H2O2 <=>  HO2- + H+}\,. \label{dissocciation}
\end{eqnarray}
Since \ce{H2O2} becomes depleted on the Pt surface, the ions in Eq.~\ref{dissocciation} will also be depleted there. The particle could then be propelled in the resulting ion gradient by ionic self-diffusiophoresis. This mechanism is identical to that of neutral self-diffusiophoresis, except that the species interacting with the particle surface to generate phoretic flow are now free ions. Such a mechanism would account naturally for the reduction of speed with salt concentration, since the mobility of charged particles in a salt gradient should scale approximately inversely with salt concentration~\cite{anderson89}. This mechanism allows, too, for direction reversals as a function of surface charge, since ionic gradients can also generate electric fields acting with or against the normal ionic diffusiophoresis~\cite{anderson89}. 

Unlike neutral diffusiophoresis, ionic diffusiophoresis is well understood theoretically, because the Coulomb interaction is fundamentally simple and long ranged (compared to, e.g., VdW or hydrophobic interactions). Thus, we can test the hypothesis of ionic diffusiophoresis quantitatively. In Appendix II, we calculate the predicted speeds in 10 $\%$ \ce{H2O2}, taking into account literature values for the disocciation of~\ce{H2O2} and the measured reaction rates. We find that, to match the observed speed, the particles would have to have a mean $\zeta$ potential of order 400 mV. This is large, but possible: for highly charged polystyrene particles $\zeta\sim$ 200 mV has been reported~\cite{midmore96}. 

A more stringent test of the ionic diffusiophoresis hypothesis comes from the NaOH data, Fig.~\ref{spiral}. NaOH dissocciates almost completely in water, removing protons from solution via \ce{OH- + H+ <=>  H2O}, hence reducing the proton gradient across the particle. We would expect then, that NaOH significantly reduces particle speed, even more than NaCl. This speed reduction does not occur. Quantitatively, we calculate in Appendix II that the ionic concentration gradient at high [NaOH] is insufficient to account for the observed propulsion by approximately 2 orders of magnitude, unless there exist surface potentials of order $\zeta\sim$ 30 V, which are impossible to achieve. Thus, we conclude that ionic dissociation alone cannot generate enough ions to drive propulsion at the observed speeds.

\begin{figure}
\centering
\hspace*{1cm}\includegraphics[width=10 cm, bb = 150 575 550 750]{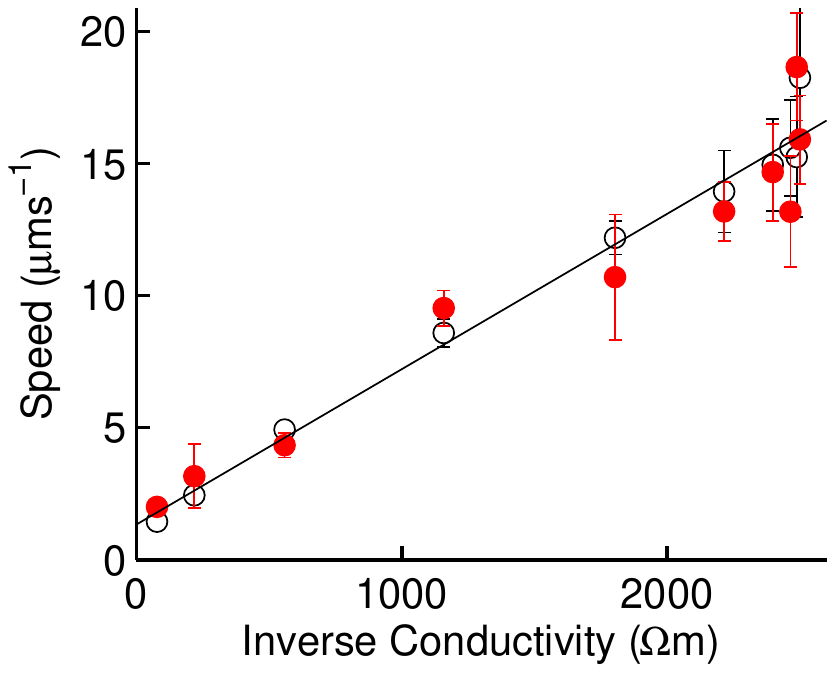}
\caption{Speed against inverse solution conductivity for  KBr ( \tikzcircle{2pt}) and NaCl($\circ$), from Fig.~\ref{All figures plot}. The solid line is a linear fit to the data, described in the text. }
\label{conductivity}
\end{figure}

\section{Self-Electrophoresis} \label{sec:electro}

We now turn to propose a tentative mechanism of self propulsion for Pt-PS Janus particles in \ce{H2O2} that is consistent with all of our data. We stress that none of our data can be deemed to prove this hypothesis. This would require separate, lengthy investigations. However, our proposal is consistent with all the data presented here, and offers simple explanations for much of it.

Like self-ionic diffusiophoresis, our proposed mechanism relies on ions. It is likely that the decomposition of \ce{H2O2} does not proceed in a single step, but involves ionic intermediates, even on a monometallic Pt surface~\cite{katsounaros12, sitta14, climent06, ono77, weiss52}. In particular, \ce{H2O2} decomposition on Pt may involve separate oxidation and reduction half-reactions, releasing and consuming ions from solution. This is known to occur in bimetallic swimmers. For example, in a Au-Pt swimmer in \ce{H2O2}, the preferred half reactions on each metal are~\cite{moran11, paxton06}:
\begin{subequations}
\begin{eqnarray}
	\ce{H2O2 + 2e- + 2H+  <=>  2H2O} \,\, \mathrm{(reduction,~Au)} \label{reduction}\\
	\ce{H2O2 <=>  2e- + 2H+ +  O2} \,\,\,\,\,\,\, \mathrm{(oxidation,~Pt)} \label{oxidation}
\end{eqnarray}
\end{subequations}
so that a a proton current flows from Pt to Au.

On a single Pt surface it is possible that the \ce{H2O2} decomposition reaction occurs by the simple disproportionation of \ce{H2O2} into \ce{H2O} and \ce{O2} without ionic intermediates. However, there is strong evidence for at least some ionic contribution. Superoxide (\ce{O2-}) ions have been directly observed in the decomposition of \ce{H2O2} on Pt~\cite{ono77}, and electrochemical measurements made under an applied voltage where there is no net oxidation or reduction show that both the oxidation and reduction reactions above can occur, depending on the surface coverage with bound \ce{OH} or \ce{O} groups~\cite{katsounaros12, sitta14, climent06}.

Even if the reaction does produce free ionic intermediates, this will not necessarily generate a current. To generate a current requires that reduction and oxidation reactions occur at different rates on different parts of the surface. Crucially, however, it is not necessary to have two different metals to produce such an ionic current. Any asymmetry across the particle surface, which affects the relative oxidation and reduction rates, would be sufficient. In the current system, there are two definite sources of asymmetry. Firstly, there will be a gradient in \ce{O2} between the equator and pole, for purely geometric reasons, and secondly, because of the directed sputtering from above, we expect a variation in Pt thickness~\cite{campbell13}, with thicker Pt at the pole than at the equator. 

It is plausible that these asymmetries will be sufficient to bias the half-reactions above, and produce a net current. For example, \ce{O2} is a product of the oxidation half reaction above, so that a higher oxidation concentration at the pole will lead to suppression of this reaction in favour of reduction, driving a proton current from equator to pole. Similarly, the thickness may affect the relative rates of reaction through variation in roughness, or through a transition from thin-layer to bulk behaviour of the catalyst. To test for the presence of this current, one could perform equivalent electrochemical experiments to those on Pt-Au swimmers~\cite{paxton06}, measuring the current between two Pt surfaces of different thicknesses or under different \ce{O2} concentration in \ce{H2O2} solution, or infer the ionic currents from the motion of tracer beads along inhomogenous Pt surfaces, as has also been done for Au-Pt surfaces~\cite{afsharfarniya13}.

It is beyond the scope of the present work to perform such measurements. However,  assuming the existence of ionic currents as hypothesised above, we apply the theoretical framework already established for bimetallic swimmers~\cite{moran10, moran11, sabass12b} to make a number of deductions that turn out to be consistent with our data. 

In a bimetallic swimmer, the electrochemical propulsion originates from the electrophoretic motion of the swimmer in the electric field generated by the ionic current. This electric field scales as $E\sim j/K$, where $j$ is the current density, and $K$ the conductivity of the solution, which for a particle of electrophoretic mobility $\mu_E =\zeta\epsilon/\eta$, gives a scaling relationship for the speed $v=\mu_EE$ of~\cite{paxton06}:
\begin{eqnarray}
	v \sim \frac{j\zeta\epsilon}{\eta K}\, , \label{scaling relation}
\end{eqnarray}
where the prefactor will depend on geometric details of the current flow. 

This predicted inverse scaling of propulsion speed to solution conductivity approximately corresponds to the inverse scaling with salt observed previously~\cite{palacci13}. Here, we correlate speed directly with conductivity. Fig.~\ref{conductivity} shows the speed obtained in the presence of KBr and NaCl replotted against inverse conductivity $1/K$. This data is well described by a linear relationship: $v=v_0 + m/K$, with $K=K_0+K_sc_s$. $K_s=12.4~\mathrm{Sm^{-1}M^{-1}}$ is the molar conductivity~\cite{henry94}, and $c_s$ the concentration, of added salt. $K_0=\mathrm{3.9\pm 0.1\times10^{-4}~\Omega^{-1} m^{-1}}$ is the measured conductivity of 10\% \ce{H2O2}. We find $v_0=1.3\mathrm{~\mu ms^{-1}}$ and $m=6\times10^{-9}\mathrm{~\Omega^{-1}s^{-1}}$. The finite residual speed $v_0$ at high conductivity could imply a small contribution from neutral diffusiophoresis. This could be tested by measuring particle speed at higher salt concentration. The variation of $\zeta$ and $j$ with salt concentration have not been taken into account, and these may also account for the offset. 

On the other hand, it is not obvious why, under this mechanism, NaOH should have little effect on the propulsion speed, since NaOH causes a similar increase in conductivity to NaCl. However, pH has a strong effect upon electrochemical oxidation and reduction of \ce{H2O2}~\cite{mckee69, hall00, venkatachalapathy99}, so NaOH can be expected to modify the current $j$, potentially compensating for this increased conductivity. 

The scaling relation Eq.~\ref{scaling relation} also predicts a reversal in propulsion direction with a reversal in surface charge (through $\zeta$), which may explain the reversal effect of CTAB, given the known ability of this surfactant to reverse the surface charge of at least polystyrene surfaces~\cite{ottewill}. However, the reverse motion of Janus particles sometimes seen with thicker Pt coatings is not predicted under this scheme. In bimetallic swimmers, the explanation for the variation of propulsion direction with different combinations of metals~\cite{wang06} is reversal of the ionic current $j$. This is also the simplest explanation for the reversal of direction with Pt thickness, and in this case, it would be expected that CTAB would still reverse the direction of motion, and salt reduce the propulsion speed, as we observe. We have no reason to predict such a current reversal with increased Pt thickness. However, in sputtered Au films, a peak in surface roughness at around 10 nm thickness has been observed~\cite{kemmenoe83}, and if the current were due to a gradient in roughness, this would naturally explain a current reversal. In addition, modifying the surface structure of Pt-Au micropumps can reverse the direction of the ionic current there~\cite{afsharfarniya13}. Sensitivity of the direction of the ionic current to the details of the catalytic surface could in general explain the variability of propulsion direction observed in the literature for catalytic Janus swimmers~\cite{ebbens12,palacci13,valadares10}.

Importantly, we can use Eq.~\ref{scaling relation} to show that the production of a current of the required magnitude in our swimmers is consistent with our measured reaction rates. Using a typical value of $\zeta=30$ mV, taken from the literature for Pt-Au swimmers~\cite{paxton06}, a speed of $v=17\mathrm{~\mu ms^{-1}}$, and $K=K_0$, as above, we require a current $j\sim 0.3 \mathrm{~Cm^{-2}s^{-1}}$. The total required flux of ions through the surface, $\Gamma_{ionic}=2\pi R^2 j/e~10^7\mathrm{s^{-1}}$, which gives $\Gamma_{\rm ionic}/\Gamma_0\sim 10^{-4}$. Hence, under this model the electrochemical part of the \ce{H2O2} reaction would only need to contribute a small fraction to the overall reaction rate to produce the observed speeds.
 
\section{Conclusion}

We have performed experiments which demonstrate strong, but selective, ionic effects on the motion of Pt-PS Janus swimmers in \ce{H2O2} solutions. As well as a reduction of speed by salt, previously found in similar haematite-PS swimmers\cite{palacci13}, we have been able to reverse the direction of motion by adding sufficient amounts of the cationic surfactant CTAB. Such effects, as well as the measured reaction rate, are found to be unexplained by generic account of propulsion via the diffusiophoresis of neutral species. Although these ionic effects could be explained by a theory of ionic diffusiophoresis based on the dissociation of \ce{H2O2}, this mechanism does not provide enough ions to generate the observed propulsion speeds.

We suggest that both the overall speed of propulsion, and the ionic effects observed here, could be explained if there were an ionic current passing between the pole and equator of the Pt particle, as in bimetallic swimmers. General asymmetries, such as chemical gradients produced by the ~\ce{H2O2} decomposition, or the variation in thickness of the Pt surface, could account for this current.

Independent of our suggested mechanism, it is clear from our experimental results that the mechanism of propulsion in catalytic Janus swimmers is considerably more complex than previously thought. Such complexity is often revealed in unexpected dependencies on parameters that are, at first sight, non-relevant. Thus, for example, we have found that changing the thickness of the Pt layer from 5~nm to 10~nm can sometimes reverse the direct of propulsion. Such dependencies begin to be understandable if differential thickness is necessary to give rise to different rates of oxidation-reduction. Not understood, such dependencies generate nuisance experimental `surprises'. Once understood, they may confer exquisite levels of control over the propulsion of such swimmers.

Acknowledgement: This work was funded by EPSRC grant EP/J007404/1. We thank Joakim Stenhammar, Ramin Golestanian, Stephen Ebbens, Mike Cates and Richard Blythe for useful discussions, and Andrew Garrie for technical assistance.

\section{Appendix I:  Neutral Diffusiophoresis Speed Calculations}

In the theory of neutral diffusiophoresis~\cite{anderson89}, the velocity $\mathbf{v}$ of a particle in an external gradient of a neutral solute $\nabla c$ is calculated. The mobility $\mu$ of the particle is defined by $\mathbf{v}=\mu\nabla c$, and it is found that, in general $\mu=k_BT\alpha/\eta$, where $\alpha$ is an interaction parameter (called $LK^*$ in the original derivation~\cite{anderson89}), which scales as length squared, and is defined as:
\begin{align}
	\alpha\,=\,\int_0^{\infty} y\left[\exp{\left(\frac{-\phi}{k_BT}\right)}-1\right] dy \,,  \label{LK}
\end{align}
where $\phi$ is the potential interaction between a solute molecule and the particle surface, and $y$ is the distance from the particle surface. Here, we modify this definition to take account of a shear plane $y_s$, which does not coincide with the particle surface. This just involves redefining the particle surface to lie a distance $y_s$ on the solution side of the liquid/particle interface:
\begin{align}
	\alpha^*\,=\,\int_{y_s}^{\infty} \left(y-y_s\right) \left[\exp{\left(\frac{-\phi}{k_BT}\right)}-1\right] dy \,,  \label{definitionAlpha}
\end{align}
In self-diffusiophoresis, one must consider the interaction of several species with the particle surface, and so the effective interaction parameter for the whole reaction is~\cite{ebbens12}):
\begin{align}
	\alpha_{\rm eff}\,=\, \frac{D_\ce{H2O2}}{2D_{O2}}\alpha^*_\ce{O2}-\alpha^*_\ce{H2O2}\,,  \label{Definition}
\end{align}
where this parameter determines the propulsion speed through Eq.~\ref{neutral speed}. An extra factor of $1/2$ in the $\ce{O2}$ term is included compared to the original definition~\cite{ebbens12} (where $\alpha_{\rm eff}$ is called $\lambda_{\rm eff}^2$) to account for the fact that only 1/2 mole of \ce{O2} is produced per mole \ce{H2O2} consumed. The \ce{O2} diffusivity is $D_\ce{O2}=3.4\times 10^{-5}\mathrm{~m^2s^{-1}}$~\cite{hung72}.

For the VdW interaction, the far field interaction potential between a molecule and a surface is given by~\cite{israelachvili11}:
\begin{eqnarray}
	\phi\,=\,\frac{-Ab^3}{3y^3} \,,  \label{VdW}
\end{eqnarray}
where $A$ is the Hamaker constant for bulk interaction of the molecule with the surface across the solvent, in this case water, and $b$ is the radius of the molecule. Inserting this potential into Eq.~\ref{definitionAlpha}, and approximating for low energy, gives:
\begin{eqnarray}
	\alpha^*\,\sim\,b^2\frac{Ab}{6k_BTy_s} \,,  \label{VdW2}
\end{eqnarray}

We assume that the slip layer begins at the edge of the Stern layer, and consists of one water molecule diameter, which we take to be $y_s=0.25$ nm~\cite{israelachvili83}. The radius of an oxygen molecule is $b=0.18$ nm~\cite{israelachvili11}, and we assume that the radius of \ce{H2O2} is similar. We do not know of any determination of the Hamaker constant of oxygen interacting with Pt in water. However, we can estimate it using the approximate formula~\cite{israelachvili11} for a dielectric (\ce{O2}) interacting through a dielectric medium (\ce{H2O}) with a metal (Pt):
\begin{eqnarray}
	A\,\sim\,\frac{3}{8\sqrt{2}}\left(\frac{n_1^2-n_3^2}{n_1^2+n_3^2}\right)\frac{h\nu_2\sqrt{\nu_1\nu_3}}{\sqrt{\nu_1\nu_3}+\nu_2/\sqrt{n_1^2+n_3^2}}
\,,  \label{VdW3}
\end{eqnarray}
where $h$ is Planck's constant, and the subscripts refer to the molecule (1), the metal (2) and the medium (3), $n$ is refractive index, and $\nu$ the fundamental adsorption frequency of the dielectrics, and the plasma frequency of the metal. There appears to be a sign error in the textbook verstion of this equation~\cite{israelachvili11}, which would give a complex result, and which we have corrected by reference to the original derivation~\cite{lipkin97}. For \ce{H2O}, we use $n_3=1.333$, $\nu_3=3\times 10^{15}$~Hz~\cite{israelachvili11}. For Pt, we use $\nu_2=1.244\times 10^{15}$~Hz~\cite{murata10}. For \ce{O2}, we use $n_1=1.2242$ for liquid oxygen~\cite{johns37}. We estimate $\nu_1=1.2\times 10^{15}$ Hz from the adsorption edge of \ce{O2} in \ce{H2O}~\cite{wu11}. Using these values, we calculate $A=-1.5\times 10^{-20}$ J, which is reasonable, since it falls between the values expected for metals interacting across water, and those for hydrocarbons interacting across water. The value of $A$ is insensitive to the precise adsorption frequencies chosen. Note that this interaction is repulsive.

Inserting this value into Eq.~\ref{VdW2}, we obtain $\alpha^*_\ce{O2}=-0.015$ nm$^2$. Repeating the calculation for \ce{H2O2}, we use $n_1=1.407$~\cite{giguere49}, and estimate $\nu_1=0.9\times 10^{15}$ Hz from the adsorption edge of \ce{H2O2} in \ce{H2O}~\cite{lin78}. This gives $A=0.9\times 10^{-20}$ J and $\alpha^*_\ce{H2O2}=0.0084$ nm$^2$. For the total interaction, Eq.~\ref{Definition} gives $\alpha_{\rm eff}=0.01$ nm$^2$, which, from Eq.~\ref{neutral speed} gives a predicted speed of $0.2~\mathrm{\mu ms^{-1}}$. Hence, under this reasonable assumption of a bound layer of water molecules, or similar, we find that the VdW interaction is insufficient to propel the particle at the observed speeds by a factor of approximately 80. Note, however, that the predicted propulsion speed strongly depends on the nature of slip on this surface. 

\section{Appendix II: Ionic Diffusiophoresis Speed Calculations}

Here we calculate the ionic self-diffusiophoretic motion of a Janus particle consuming hydrogen peroxide on one face when there is a local equilibrium between the hydrogen peroxide and other species in solution. We show that adding NaOH to the solution lowers the predicted speed from this mechanism to the extent that it cannot account for the observed propulsion speed of our Pt-PS Janus particles.

We first consider the general case of a fixed, charged, material surface in contact with a fluid in which is dissolved a neutral molecule (here \ce{H2O2}) with concentration field $c_0(\mathbf{r})$, and diffusivity $D_0$, which is in local equilibrium with a number $n$ of ionic species with concentration $\{c_i(\mathbf{r})\}$, valency $\{z_i\}$ (limited here to $\{z_i\}=\pm 1$), and diffusivity $\{D_i\}$ for $i=1:n$. The bulk concentrations are given by $c_{0,\infty}$, and $\{c_{i,\infty}\}$ respectively. 

We assume there is no direct interaction between the neutral molecule and the surface. The ions, however, interact with the charged surface through the Coulomb interaction, so that a gradient of ions along the surface produces a tangential flow. In addition, when the diffusion rates of ions are unequal, this generates an electric field to maintain bulk charge neutrality, and this electric field also produces tangential flow. Together, these effects are called ionic diffusiophoresis~\cite{anderson84,anderson89}. The net result is flow in a thin interfacial layer (the Debye layer), at the outer edge of which the flow rate is given by an effective slip velocity $\mathbf{v_s}$, where:
\begin{eqnarray}
	 \mathbf{v_s} \,=\, -\left( \mathbb{I} - \mathbf{nn}\right)\cdot \left(\mu_D \mathbf{D}  + \mu_E \mathbf{E}       \right)                           \,. \label{slip velocity}
\end{eqnarray}
Here $\mathbb{I}$ is the identity matrix, and $\mathbf{n}$ is the normal to the surface, so that the operator $\left( \mathbb{I} - \mathbf{nn}\right)\cdot$ returns only those components of a vector which are tangential to the surface. $\mu_D$ and $\mu_E$ are the diffusiophoretic and electrophoretic mobilities of the surface, which are properties of the surface that may vary with position. We will give expressions for these mobilities shortly. We define the ionic gradient $\mathbf{D}$ as:
\begin{eqnarray}
	\mathbf{D}\,=\,\frac{\sum_i\nabla c_{i}}{\sum_i{z_i^2c_i}}                         \,, \label{nabla c ion}
\end{eqnarray}
where we have extended the treatment for $n=2$ in~\cite{anderson89} to consider general $n$. The electric field $\mathbf{E}$, which arises from the differential diffusivity of the various ions, is, by a similar extension:
\begin{eqnarray}
	 \mathbf{E} = \frac{k_BT}{e}\frac{\sum_iz_iD_i\nabla c_i}{\sum_iz_i^2D_ic_i}       \,, \label{electricField}
\end{eqnarray}
where $e$, $k_B$ and $T$ are respectively the charge on the proton, Boltzmann's constant, and temperature. In the limit of small ionic gradients, we can approximate the concentration fields in the denominator of $\mathbf{D}$ and $\mathbf{E}$ with their bulk values, $\{c_{i,\infty}\}$. In addition, if the ionic concentrations are in equilibrium with $c_0$, the gradients are given by:
\begin{eqnarray}
	 \nabla c_i = \frac{\partial c_i}{\partial c_0} \nabla c_0       \,. \label{ionic gradients}
\end{eqnarray}
One can then express the slip velocity as a function of the gradient of the neutral molecule $\nabla c_0$, and a mobility which only depends on local properties of the surface combined with ionic concentrations in the bulk, and their equlibria with respect to the neutral molecule. We write:
\begin{eqnarray}
	 \mathbf{v_s} \,=\, \frac{-1}{c_{0,\infty}}\left(\mu_D \tilde{D}  + \mu_E\frac{k_BT}{e} \tilde{E}       \right)\left( \mathbb{I} - \mathbf{nn}\right)\cdot \nabla c_0                           \,, \label{slip velocity neutral}
\end{eqnarray}
where the dimensionless parameters $\tilde{D}$ and $\tilde{E}$ are:
\begin{eqnarray}
	\tilde{D}\,=\,\frac{c_{0,\infty}}{\sum_i{z_i^2c_{i,\infty}}} \sum_i\frac{\partial c_i}{\partial c_0}      \,, \label{D tilde}
\end{eqnarray}
\begin{eqnarray}
	 \tilde{E} = \frac{c_{0,\infty}}{\sum_iz_i^2D_ic_{i,\infty}}\sum_iz_iD_i\frac{\partial c_i}{\partial c_0}       \,. \label{E tilde}
\end{eqnarray}
The total mobility $\mu$ of the surface with respect to the concentration gradient $\nabla c_0$ is defined by:
\begin{eqnarray}
	 \mathbf{v_s} \,=\, -\mu \left( \mathbb{I} - \mathbf{nn}\right)\cdot \nabla c_0                           \,, \label{mobility}
\end{eqnarray}
so that
\begin{eqnarray}
	 \mu \,=\, \frac{1}{c_{0,\infty}}\left(\mu_D \tilde{D}  + \mu_E\frac{k_BT}{e} \tilde{E}       \right)   \,. \label{total mobility}
\end{eqnarray}

\begin{figure}[t]
  \centering 
\vspace*{-1cm}  \hspace*{1cm} \includegraphics[width=10cm, bb =150 575 550 800]{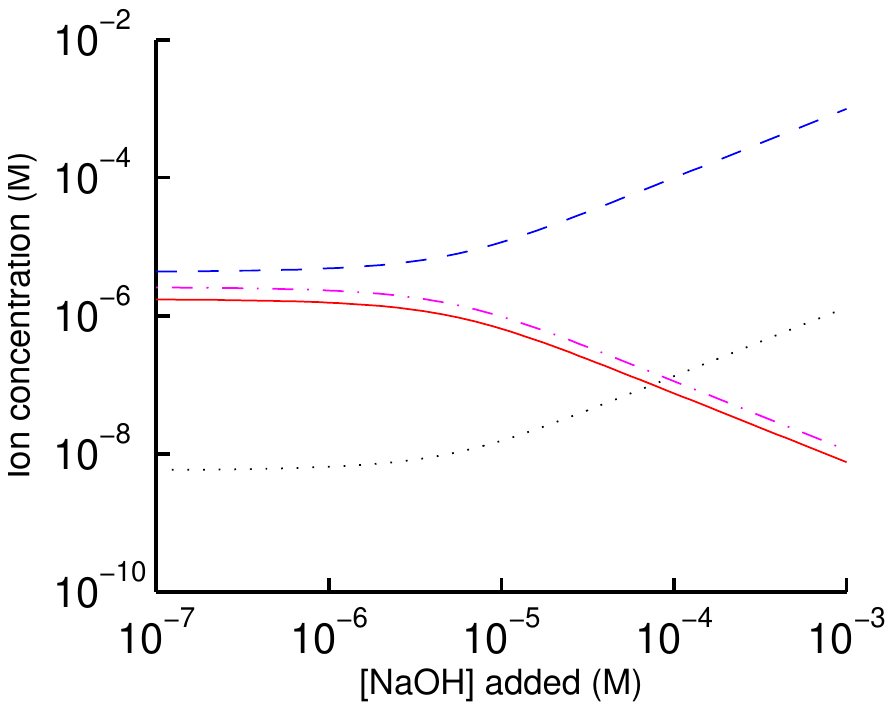}
  \caption{Ionic concentrations in $10\%~\mathrm{H_2O_2}$ as a function of [NaOH]. \ce{H+}, solid red;  \ce{HO2-}, blue dashed; \ce{OH-}, black dotted; \ce{H3O2+}, magenta dot-dashed.}
\label{concentrations}
\end{figure}

\begin{figure}[t]
  \centering 
\vspace*{-1cm}   \hspace*{1cm} \includegraphics[width=10cm, bb =150 575 550 800]{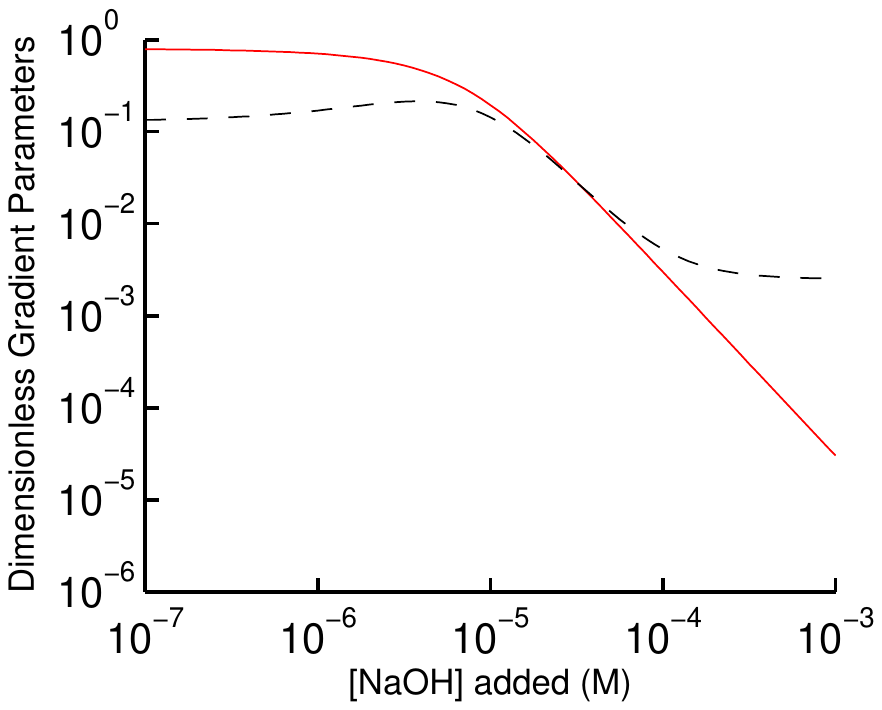}
  \caption{Calculated values of the dimensionless parameters $\tilde{D}$ (red solid), which measures the ion gradient giving rise to diffusiophoresis, and $\tilde{E}$ (black dashed), which measures the electric field produced by this ion gradient. For 10 $\%$ \ce{H2O2} and varying [NaOH].}
\label{dimensionless}
\end{figure}

This result (Eq.\ref{slip velocity neutral}), which is for flow over a fixed surface, can be converted via the reciprocal theorem in low Reynolds number hydrodynamics~\cite{golestanian07,stone96,happel83} to the velocity $\mathbf{v}$ of a free particle in a concentration field.  In particular~\cite{golestanian07} has shown how to calculate the speed of a particle which generates its own concentration gradient through surface reactions. In the relevant case for us, of a Janus sphere, with hemispheres labelled $j=1,2$, which consumes the neutral reactant uniformly on the catalytic ($j=1$) face, with total rate $\Gamma$, and has different mobilities, $\mu_1$ and $\mu_2$ on the two hemispheres, it has been shown that~\cite{golestanian07}:
\begin{eqnarray}
	 \mathbf{v} \,=\, \frac{\left(\mu_1+\mu_2\right)\Gamma}{16\pi D_0 R^2} \mathbf{z}  \,, \label{particle speed}
\end{eqnarray}
where $\mathbf{z}$ is the unit vector pointing towards the non-catalytic ($j=2$) pole of the particle, and $R$ is the particle radius. This model corresponds to our current experimental system, with $\Gamma$ the rate of consumption of \ce{H2O2}, and $D_0=D_\ce{H2O2}$. To determine whether ionic diffusiophoresis can explain the observed propulsion, it remains then to estimate $\mu_E$ and $\mu_D$, and to calculate the equilibrium concentrations and $\partial c_i /\partial c_0$ for the ions in equilibrium with hydrogen peroxide.

To calculate the ionic concentrations, we solve the relevant equilibrium equations. The ionic equilibrium of \ce{H2O2} is slightly more complicated than the simple disocciation discussed in the main text, since \ce{H2O2} is both deprotonated and protonated~\cite{kosinski07} in aqueous solution. The relevant reactions are hence:
\begin{eqnarray*}
	\ce{H2O <=> H+ + OH- }\,,\\ \ce{H2O2 <=> H+ + HO2- }\,,\\ \ce{H+ + H2O2 <=> H3O2+}\,,
\end{eqnarray*}
We add NaOH at concentration $[\ce{NaOH}]_0$ (between 0 mM and 1 mM). For simplicity, we make the approximation that NaOH is fully disocciated, giving $[\ce{Na+}]=[\ce{NaOH}]_0$. We also assume that \ce{H2O2} is in sufficient excess to maintain its original concentration ($[\ce{H2O2}]=[\ce{H2O2}]_0$), where $[\ce{H2O2}]_0$ is the spatially varying concentration of \ce{H2O2} determined by the flux of \ce{H2O2} into the Pt surface. We have checked that these simplifying approximations do not affect our conclusions. Under these approximations, the equations to solve are the equilibria:
\begin{eqnarray*}
	[\ce{H+}][\ce{OH-}]\,=\,k_W  \,, \\ \\
	\frac{[\ce{H+}][\ce{HO2-}]}{[\ce{H_2O_2}]}\,=\,k_D \,, \\ \\	
	\frac{[\ce{H3O2+}]}{[\ce{H+}][\ce{H2O2}]}\,=\,k_P \,,
\end{eqnarray*}
and charge conservation equation:
\begin{eqnarray*}
	[\ce{H+}] + [\ce{H3O2+}] + [\ce{Na+}] \,=\, [\ce{OH-}] + [\ce{HO2-}]
\end{eqnarray*}
where $k_D=2.5\times 10^{-12}$~\cite{everett53}, $k_W=10^{-14}$, and we estimate $k_P$ from the difference between measured and active pH discussed in~\cite{kosinski07}. There, the pH measured using an electrode in 10 $\%$ \ce{H2O2} is $\mathrm{pH_E}=5.1$, whereas the titrated pH, $\mathrm{pH_T}=5.5$ . The difference is attributed to \ce{H3O2+} ions behaving like protons with respect to the electrode, so that $\mathrm{pH_E}=-\log_{10}{([\ce{H+}] + [\ce{H3O2+}])}$ and $\mathrm{pH_T}=-\log_{10}{([\ce{H+}])}$ only. From this we calculate $k_P=0.5$. A similar estimate from the data in~\cite{kolczynski57} yields $k_P=0.2$. We find, in any case, that the value of this equilibrium constant makes no difference to our conclusions, and only affects $\tilde{E}$ at low [NaOH].  

Solving these equations gives a quadratic equation in [\ce{H+}], from which we obtain the concentrations of all species, as well as the partial derivatives $\partial c_i/\partial[\ce{H2O2}]$. The ionic concentrations are shown in Fig.~\ref{concentrations} for 10\% \ce{H2O2} and varying [NaOH]. We do not include \ce{Na+} as its concentration is equal to that of the added NaOH. At high added [NaOH], the [\ce{H+}] becomes vanishingly small, which limits the total ionic gradient. 

Next, we calculate the parameters $\tilde{D}$ and $\tilde{E}$ (Fig.~\ref{dimensionless}). At high added [NaOH], $\tilde{D}$ falls by 4 orders of magnitude. $\tilde{E}$ does not fall so dramatically because although there is a very small net ionic gradient, the different diffusivities of the ions mean that there is still a moderate electric field. We use $D_\ce{H+}=9.3\times10^{-9}~\mathrm{m^2 s^{-1}}$, $D_\ce{Na+}=1.3\times10^{-9}~\mathrm{m^2 s^{-1}}$, $D_\ce{OH-}=5.1\times10^{-9}~\mathrm{m^2 s^{-1}}$~\cite{henry94}, and $D_\ce{HO2-}=0.9\times10^{-9}~\mathrm{m^2 s^{-1}}$~\cite{vandenbrink84}. In the absence of experimental values, we assume that $D_\ce{H3O2+}= D_\ce{HO2-}$, but setting $D_\ce{H3O2+}= D_\ce{H+}$ again makes no difference to our conclusions, and only affects $\tilde{E}$ at low added [NaOH]. We have also repeated these calculations taking into account typical atmospheric concentrations of \ce{CO2} (400 ppm), which forms \ce{H2CO3} in aqueous solution. This did not affect our results.

\begin{figure}[t]
  \centering 
\vspace*{-1cm}  \hspace*{1cm} \includegraphics[width=10cm, bb =150 575 550 800]{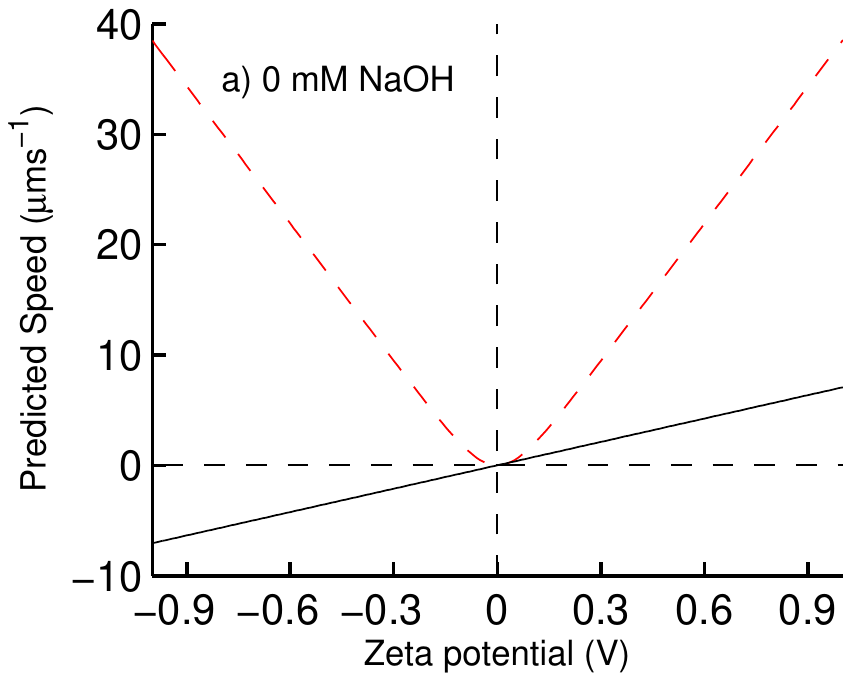}
\vspace*{-1cm}   \hspace*{1cm} \includegraphics[width=10cm, bb =150 550 550 800]{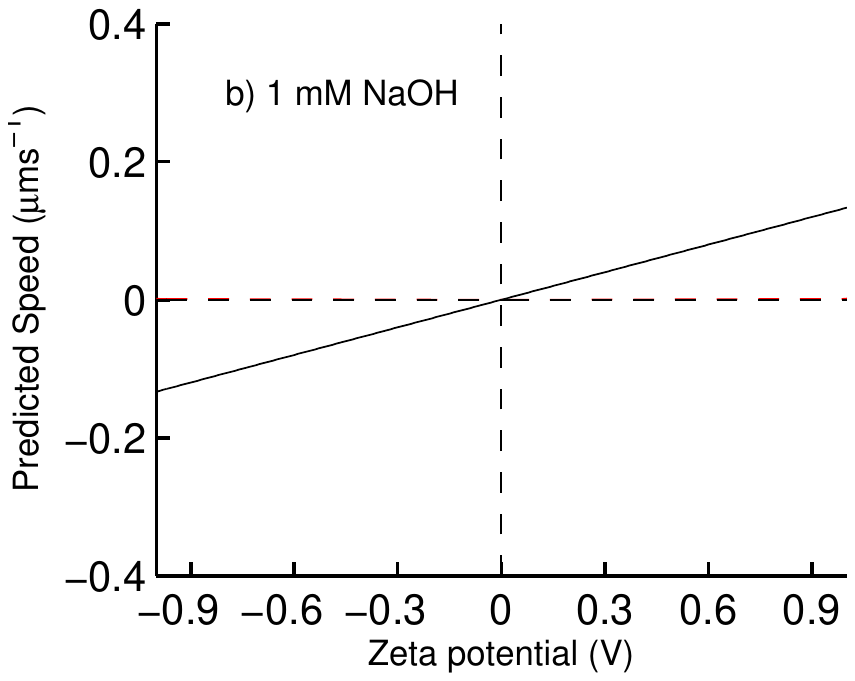}
\vspace*{.3cm}
  \caption{Predicted speed in 10$\%$ \ce{H2O2} and a) 0 mM, b) 1 mM NaOH from Eq.~\ref{particle speed}, with $\Gamma=8\times10^{10}\mathrm{~s^{-1}}$. The solid black line shows the electrophoretic, and dashed red diffusiophoretic contributions. Positive motion is towards the uncoated face of the particle. In b) the diffusiophoretic contribution lies on the x-axis.}
\label{speed}
\end{figure}

In the limit of small Debye lengths, both electrophoretic and diffusiophoretic mobilities have analytical solutions given by~\cite{anderson84}:
\begin{eqnarray}
	\mu_D\,=\, \frac{4\epsilon_R\epsilon_0}{\eta}\left(\frac{k_BT}{e}\right)^2\log{\left(\cosh{\left(\frac{\zeta e}{4k_BT}\right)}\right)} \,, \label{diffusiophoretic}
\end{eqnarray}
\begin{eqnarray}
	\mu_E\,=\, \frac{\zeta\epsilon_0\epsilon_R}{\eta} \,, \label{electrophoretic}
\end{eqnarray}
where $\epsilon_0$ is the vacuum permittivity, $\epsilon_R$ is the relative permittivity of the solution, and $\eta$ is the solution viscosity. We approximate $10\%$ \ce{H2O2} as having similar properties to water, so that $\epsilon_R=78$, and $\eta=10^{-3}$ Pa~s. The particle zeta potential, $\zeta$, may be different on the two halves of the particle, and itself depends on ionic concentration and surface charge density. Outside the small Debye length limit, Eq.~\ref{diffusiophoretic} and Eq.~\ref{electrophoretic} no longer apply, and both $\mu_D$~\cite{prieve87} and $\mu_E$~\cite{midmore96} must be calculated numerically. They are found to become non-monotonic, and have maximum values at typical values of $\zeta$ of order 100 mV, depending on the thickness of the Debye layer. The peak in electrophoretic mobility in Fig.3b demonstrates that we are outside the low Debye length regime~\cite{midmore96}. 

We therefore take Eq.~\ref{diffusiophoretic} and Eq.~\ref{electrophoretic} to represent upper bounds on these mobilities. In this case, we cannot draw conclusions from any cancellation of electrophoretic mobility and diffusiophoretic mobility acting in opposition, but consider each individually. From these upper bounds, the predicted contributions to the speed from diffusiophoresis and electrophoresis at 10$\%$ \ce{H2O2} and 0 or 1 mM NaOH are shown in Fig.~\ref{speed}. At 0 mM NaOH, the observed speed ($\geq$15~$\mathrm{\mu ms^{-1}}$, Fig.2) is achievable with an average $\zeta$ potential of approximately -400 mV (from diffusiophoresis alone), which is rather large, but feasible. At 1 mM NaOH, however, the observed propulsion speed ($\geq$4~$\mathrm{\mu ms^{-1}}$, Fig. 2) would require $|\zeta|>30$ V, which is not practically achievable, and may be physically impossible outside the small Debye length limit because of the maximum in the mobility mentioned above. In addition, for a negatively charged particle ($\zeta<0$), this model predicts motion in the wrong direction, i.e. towards the Pt face at high NaOH. These objections to ionic diffusiophoresis apply also at lower NaOH concentrations. The $\zeta$ required to explain the observed propulsion (Fig.2) is over 1 V for [NaOH] above about 0.03~mM. 

We note that the derivation of electrophoretic and diffusiophoretic mobility in~\cite{anderson89, prieve87, midmore96} assumes a strict hydrodynamic no-slip condition at the particle surface. It has been shown in~\cite{ajdari06} that a finite slip velocity on the particle surface, which is not to be confused with $\mathbf{v_s}$, the effective slip velocity at the edge of the interfacial layer, could generate much larger mobilities, scaling mobility by factors of order $1+b/\lambda_D$ for charged diffusiophoresis, where $b$ is the distance inside the solid surface at which the flow speed would extrapolate to zero. However, $\lambda_D>10$ nm for salt concentrations below 1 mM, and we do not expect $b$ larger than 20-30 nm~\cite{ajdari06, cottinbizonne05},  so this effect is also probably insufficient to explain the observed propulsion speeds at high [NaOH]. 

\footnotesize{
\bibliography{PROPULSIONREFERENCELIST}	
\bibliographystyle{rsc} 
}

\end{document}